%% file: towards-coexecution-commodity-time-constrained.tex
\documentclass[conference]{IEEEtran}
\usepackage{makeidx}  % allows for index generation

\usepackage{multirow}
\usepackage{varwidth} %for the varwidth minipage environment
\usepackage{moresize}%\ssmall
\usepackage{ifthen}
\usepackage[hyphens]{url}

\usepackage{tikz}
\usetikzlibrary{calc,arrows}

\usepackage{xcolor}
\usepackage{colortbl}

\usepackage{tabularx}
\usepackage{array}

\usepackage{ulem}
\definecolor{light-gray}{gray}{0.95}
\definecolor{MyDarkGreen}{rgb}{0,0.8,0.0}
\definecolor{MyDarkBlue}{rgb}{0,0,0.8}
\definecolor{MyDarkRed}{rgb}{0.6,0,0.0}
\definecolor{MyOrange}{rgb}{0.6,0,0.0}
\usepackage{multicol} % added package
\usepackage{relsize}
\usepackage{xspace}
\PassOptionsToPackage{hyphens}{url}\usepackage{hyperref}
\newcommand\Cpp{C\nolinebreak[4]\hspace{-.05em}\raisebox{.4ex}{\relsize{-3}{\textbf{++}}}}
\usepackage{float} % using minipage with minted, you can use listing[H]
\usepackage{dblfloatfix}    % To enable figures at the bottom of page

\usepackage{fixltx2e}
\usepackage{amsmath}
\usepackage{amssymb}
\usepackage{mathtools}

\DeclarePairedDelimiter\floor{\lfloor}{\rfloor}

\ifpdftex
\usepackage[T1]{fontenc}
\usepackage[nomath]{lmodern}
\else
\usepackage{fontspec}
\fi
\usepackage[frozencache,cachedir=cache]{minted}

\usepackage{fancyhdr}
\fancypagestyle{pprintTitle}{%
  \chead{Accepted in Int. Conf. on High Perf. Computing and Simulation (HPCS - WEHA) \url{https://ieeexplore.ieee.org/document/9188188}}%
\lfoot{}\cfoot{}\rfoot{}

}
\usepackage[most,minted]{tcolorbox}
\tcbuselibrary{minted,skins}
\usepackage{etoolbox}
\newtcblisting{snippet}[2][]{
    listing engine=minted,
    colback=light-gray,
    colframe=white,
    listing only,
    minted style=default,%colorful
    minted language=#2,
    minted options={linenos=true,numbersep=1mm,fontsize=\small,#1},
    left=14pt,
    enhanced,
    overlay={\begin{tcbclipinterior}\fill[black!15] (frame.south west)
            rectangle ([xshift=4.5mm]frame.north west);\end{tcbclipinterior}}
}
\renewcommand{\arraystretch}{1.1}

\usepackage{hyperref} %% RNOZ

\hypersetup{
bookmarks=true,         % show bookmarks bar?
unicode=false,          % non-Latin characters in Acrobat’s bookmarks
pdftitle={Towards Co-execution on Commodity Heterogeneous Systems: Optimizations for Time-Constrained Scenarios},
pdfauthor={Raul Nozal, Jose Luis Bosque and Ramon Beivide},
pdfkeywords={
Heterogeneous Computing,
Co-execution, 
Commodity Systems,
OpenCL,
Load Balancing,
CPU,
GPU,
Scheduling
},
pdfsubject={System Software},
pdfproducer={LaTeX,TeX},
pdfcreator={xelatex}
}

\begin{document}

% \textbf{XXX: when Usability reaches the Performance of Heterogeneous Computing}

%\tableofcontents
%
%% SBAC:
%% \mainmatter              % start of the contributions
%
% Brainstorming will come: Load balancing in commodity infrastructures: is it worth it?
%\title{Towards Co-execution on Commodity Heterogeneous Systems: Optimizations under time constrained offloading}
\title{Towards Co-execution on Commodity Heterogeneous Systems: Optimizations for Time-Constrained Scenarios}

%Towards Co-execution on Commodity Heterogeneous Systems: Time-Constrained Offloading Optimizations
%Towards Co-execution on Commodity Heterogeneous Systems: time-constrained offloading optimizations
%Towards Co-execution on Commodity Heterogeneous Systems: Time-Constrained Scheduling Optimizations
%Towards Co-execution on Commodity Heterogeneous Systems: time-constrained scheduling optimizations
% Towards Co-execution on Commodity Heterogeneous Systems: Optimizations for Time-constrained Scenarios
%\title{Towards Co-execution on Commodity Heterogeneous Systems}
%\title{Towards Co-execution on Commodity Heterogeneous Systems}

\author{
  \IEEEauthorblockN{Ra\'ul Nozal, Jose Luis Bosque and Ram\'on Beivide}
  \IEEEauthorblockA{Computer Science and Electronics Department, Universidad de Cantabria, Spain\\
    \{raul.nozal,joseluis.bosque,ramon.beivide\}@unican.es}
}

\maketitle\thispagestyle{pprintTitle}             % typeset the title of the contribution

%% Page Numbers

%\ifthenelse{\equal{\withPageNumbers}{yes}}{%
%\thispagestyle{plain}
%\pagestyle{plain}
%}{}

%This paper presents EngineCL, a new OpenCL-based runtime system that significantly improves the usability of the heterogeneous systems with slightly loss of performance.

% It is developed in C++ and
% NOZAL: TODO faltaría algo del balanceo

%\thispagestyle{pprintTitle}
%\pagestyle{pprintTitle}
\input{sections/abstract.tex}

% TODO: en la version revista usar un % geomean, pero también para los dispositivos no CPU, ya que es mucho menor.

%% SBAC:
%% \begin{IEEEkeywords}
%% Heterogeneous Computing,
%% Usability,
%% Performance portability,
%% OpenCL,
%% Parallel Programming,
%% Scheduling,
%% Load balancing,
%% Productivity,
%% API
%% \end{IEEEkeywords}

%% \vspace{-2mm}

\input{sections/introduction.tex}

\pagestyle{plain}
\input{sections/design.tex}

%\input{sections/api.tex}

\input{sections/methodology.tex}

\input{sections/results.tex}

\input{sections/related-work.tex}

\input{sections/conclusions.tex}

\section*{Acknowledgement}
This work has been supported by the the Spanish Ministry of Education (FPU16/ 03299 grant), the Spanish Science and Technology Commission (TIN2016-76635-C2-2-R), the  European Union's Horizon 2020 research and innovation programme and  HiPEAC Network of Excellence (Mont-Blanc project under grant 671697).

% Network  of  Excellence  and  the  European  Union's  Horizon 2020 research and innovation programme (Mont-Blanc project under grant agreement No 671697).
% the European Research Council (G.A. No 321253) and the European HiPEAC Network of Excellence.
% The Mont-Blanc project has received funding from the European Union’s Horizon 2020 research and innovation programme under grant agreement No 671697.

%% \vspace{-4mm}
\bibliographystyle{splncs03}

%% \bibliography{maat,power,europar2018,europar18,usability}

%% \clearpage
%% \addtocmark[2]{Author Index} % additional numbered TOC entry
%% \renewcommand{\indexname}{Author Index}
%% \printindex
%% \clearpage
%% \addtocmark[2]{Subject Index} % additional numbered TOC entry
%% \markboth{Subject Index}{Subject Index}
%% \renewcommand{\indexname}{Subject Index}
%\input{subjidx.ind}

\end{document}

%% file: sections/abstract.tex
\begin{abstract}
Heterogeneous systems are present from powerful supercomputers, to mobile devices, including desktop computers, thanks to their excellent performance and energy consumption.
The ubiquity of these architectures in both desktop systems and medium-sized service servers allow enough variability to exploit a wide range of problems, such as multimedia workloads, video encoding, image filtering and inference in machine learning.
Due to the heterogeneity, some efforts have been done to reduce the programming effort and preserve performance portability, but these systems include a set of challenges. The context in which applications offload the workload along with the management overheads introduced when doing co-execution, penalize the performance gains under time-constrained scenarios.
Therefore, this paper proposes optimizations for the EngineCL runtime to reduce the penalization when co-executing in commodity systems, as well as algorithmic improvements when load balancing. An exhaustive experimental evaluation is performed, showing optimization improvements of 7.5\% and 17.4\% for binary and ROI-based offloading modes, respectively. Thanks to all the optimizations, the new load balancing algorithm is always the most efficient scheduling configuration, achieving an average efficiency of 0.84 under a pessimistic scenario.

\end{abstract}
% Sistema en el que estamos (commodity) y problemas como los descritos (video encoding, etc)
% Presentan retos a la co-ejecución
% En este paper se proponen optimizaciones en algoritmica y rutime
% Evaluación experimental
% conseguir ese 0.84 de eficiencia 3%, se situa sobre el resto. Con las optimizaciones efectuadas siempre se es rentable.

\begin{IEEEkeywords}
Heterogeneous Computing,
Co-execution, 
Commodity Systems,
OpenCL,
Load Balancing,
CPU,
GPU,
Scheduling.
\end{IEEEkeywords}

%% file: sections/introduction.tex
\section{Introduction}
\label{sec:intro}

Heterogeneous systems made up of a general purpose CPU and a set of hardware accelerators are increasingly present in a great number of compute devices. Thus, we can find them from powerful computing nodes in great supercomputers, to mobile devices, including in this range commodity computers. Desktop computers usually have an integrated heterogeneous systems, composed of CPU cores, together with GPU compute units in a single chip. Along with them, it is common to find discrete GPUs.
%that, without reaching the performance achieved by high-end GPUs, offer excellent performance and energy efficiency.
% última frase se podría quitar (without...)
% Meter el de FPGAs

The ubiquity of these devices allows enough variability of heterogeneous systems that, together with their architectures, make it easier to exploit a wide range of problems, such as multimedia workloads, video encoding, image filtering and inference in machine learning. The availability of these systems in both commodity infrastructures and medium-sized service servers facilitate a new field of work, but involves great challenges. The scenario in which these offloading functions are launched is determined by executions of hundreds of milliseconds, sometimes a few seconds, where every management operation or the minimum overhead completely penalizes the offloading to devices.

In addition, these functions are usually carried out in two operating modes. On the one hand, integrated in a main program, directly consuming the data and making only the transfer and computation through the use of other technologies such as OpenMP or OpenCL. This function is executed in parallel while the main program continues operating, such as the server managing requests or the GUI rendering charts. On the other hand, by launching this function as a process independently to the main program, serializing and exchanging the data and results.

Therefore, it is important to take into account the context in which the applications are executed, as it has been done in other cases with devices such as FPGAs \cite{Guzman:2019}, knowing that we are facing the worst possible scenario to do co-execution. Furthermore, it is necessary to exploit techniques and optimizations to allow an efficient execution that takes advantage of all the available resources.
% como se ha tenido en cuenta en otros casos con dispositivos FPGAs (JoS).
%% The scenario in which these applications are framed is determined by executions of hundreds of milliseconds, sometimes a few seconds, where a minimum overhead completely penalizes the download to devices.

A key aspect of maximizing the performance of such systems is {\em co-execution}. That is, all devices in the system cooperate to execute a single massively data-parallel task, improving its response time. The co-execution allows to extract the maximum performance of the system, as well as to optimize its energetic consumption, since all the devices contribute useful work to solve the problem, instead of remaining idle. However, the programming models currently used do not have adequate support for co-execution. For instance, OpenCL \cite{OpenCL:Gaster:2013} provides low abstraction level that forces the programmer to know the system in detail and manually do a set of operations, such as managing the host-device communication or explicitly partitioning the data among the devices.

To solve these problems and provide effortless co-execution it is necessary to overcome two important challenges: {\em abstraction} and {\em load balancing}. Abstraction means that the programmer should always work in the same way, regardless of the system in which his application is executed. Therefore, all system-dependent tasks should be hidden from him. Load balancing refers to a distribution of work proportional to the computational capacity of each of the devices. Thus, everyone should finish their work approximately at the same time, avoiding that none of them remain idle but consuming energy.

To overcome these problems this paper improves some features of  {\em EngineCL} \cite{EngineCL:2018} to desktop systems, composed by a multi-core CPU, an integrated GPU and a middle-range discrete GPU. EngineCL is an OpenCL-based \Cpp{} runtime API that significantly improves the usability of the heterogeneous systems without any loss of performance. It acts as a wrapper to facilitate the programming while providing flexibility and a system to evaluate scheduling algorithms. It accomplishes complex operations transparently for the programmer, such as discovery of platforms and devices, data management, load balancing and robustness throughout a set of efficient techniques. Following the Host-Device programming model, the runtime manages a single data-parallel kernel among all the devices in the heterogeneous system.

% añadir nuevas optimizaciones a EngineCL...
% pasar a segundo termino lo que está

%%JLB: Intro multiprog
%%EngineCL has been validated both in terms of usability and performance, using a system composed of three different architectures: CPU, Xeon Phi and GPU. Regarding usability, 5 metrics have been used, achieving excellent results in all of them. In terms of performance, the runtime overhead compared with OpenCL is on average around 0.46\% when using a single device.
Two types of optimizations have been researched: runtime-centered to allow using load balancing under more problem sizes, competing against the fastest device in the system, and algorithmically, proposing a new algorithm and tuning its parameters to boost the average efficiency in a wide range of program types. The runtime improvements have decreased the management and synchronization overheads due to initialization and buffer optimizations, increasing the general efficiency in 7.5\% and 17.4\% for binary and ROI-based offloading modes. Finally, thanks to the new proposals, when using all the devices in the system to solve a problem, the performance is greatly improved, increasing around 3\% on average with respect to the default HGuided algorithm, yielding an efficiency of 0.84 and a balance effectiveness of 0.97. 

% Finally, when using all the devices in the system to solve a problem thanks to the provided schedulers, the performance is greatly improved. The geometric mean of efficiencies for the optimized HGuided algorithm is 0.84 and a balance effectiveness of 0.97. The improvements due to the new optimizations is around 3\% with respect to the default HGuided algorithm.
% Agregar los resultados de optimizaciones 7.5 y 17.40%, que dan lugar a una mejora del 3% en eficiencia etc

The main contributions of this paper are the following:
\begin{itemize}
\item New EngineCL optimizations are proposed to mitigate the overhead of the initialization and finalization stages, as well as buffer management, caused during the co-execution by the use of OpenCL drivers in commodity computers.

\item An exhaustive experimental evaluation of the HGuided algorithm, searching for the best tuning of its parameters to optimize its performance in desktop systems.
\end{itemize}

%%Intro Multiprog.
%\begin{itemize}
%\item Presents EngineCL, a runtime that % notably
%  extraordinarily simplifies the programming of data-parallel application on a heterogeneous system.
%En el seguno item añadir que proporcional eq. de carga.
%\item EngineCL ensures performance portability fully exploiting heterogeneous machines.
%\item An exhaustive experimental validation, both of the usability and the performance of the runtime,
  % which allows to conclude its excellent behavior in both metrics.
%  proving its excellent behaviour in both metrics.
%\end{itemize}

%JLB: Revisar este párrafo cuando definamos la estructura definitiva del artículo.
The rest of this paper is organized as follows. Section \ref{sec:Overview} describe the design and implementation of EngineCL, along with the new HGuided algorithm. The proposed optimizations are explained in Section \ref{sec:Optimization}.
%Section \ref{sec:API} presents two examples of how to use the API. 
The methodology used for the validation is explained in Section \ref{sec:Methodology}, while the experimental results are shown in Section \ref{sec:ExperimentalResults}. Finally, Section \ref{sec:Related} explains similar works while Section \ref{sec:Conclusions}, the most important conclusions and future work are presented.

%% file: sections/design.tex
\section{EngineCL Overview}
\label{sec:Overview}

\subsection{Principles of Design and Implementation}
\label{sec:Design}

% EngineCL has been designed with many principles in mind, all around three pillars: OpenCL, Usability and Performance.
EngineCL is designed with many principles in mind, all around three pillars: OpenCL, Usability and Performance.

It is tightly coupled to OpenCL and how it works.
% The modules of the system and its relations have been defined based on the most performant and stable patterns.
The system modules and their relationships have been defined according to the most efficient and stable patterns.
Every design decision has been benchmarked and profiled to achieve the most optimal solution in every of its parts, but mainly promoting the modules related with the data management, synchronization and API abstraction.

EngineCL is designed to achieve high external usability and internal adaptability to support new runtime features when the performance is not penalized.

\begin{figure}[H]
   \centering
   \hspace{-2mm}
   \includegraphics[width=0.45\textwidth]{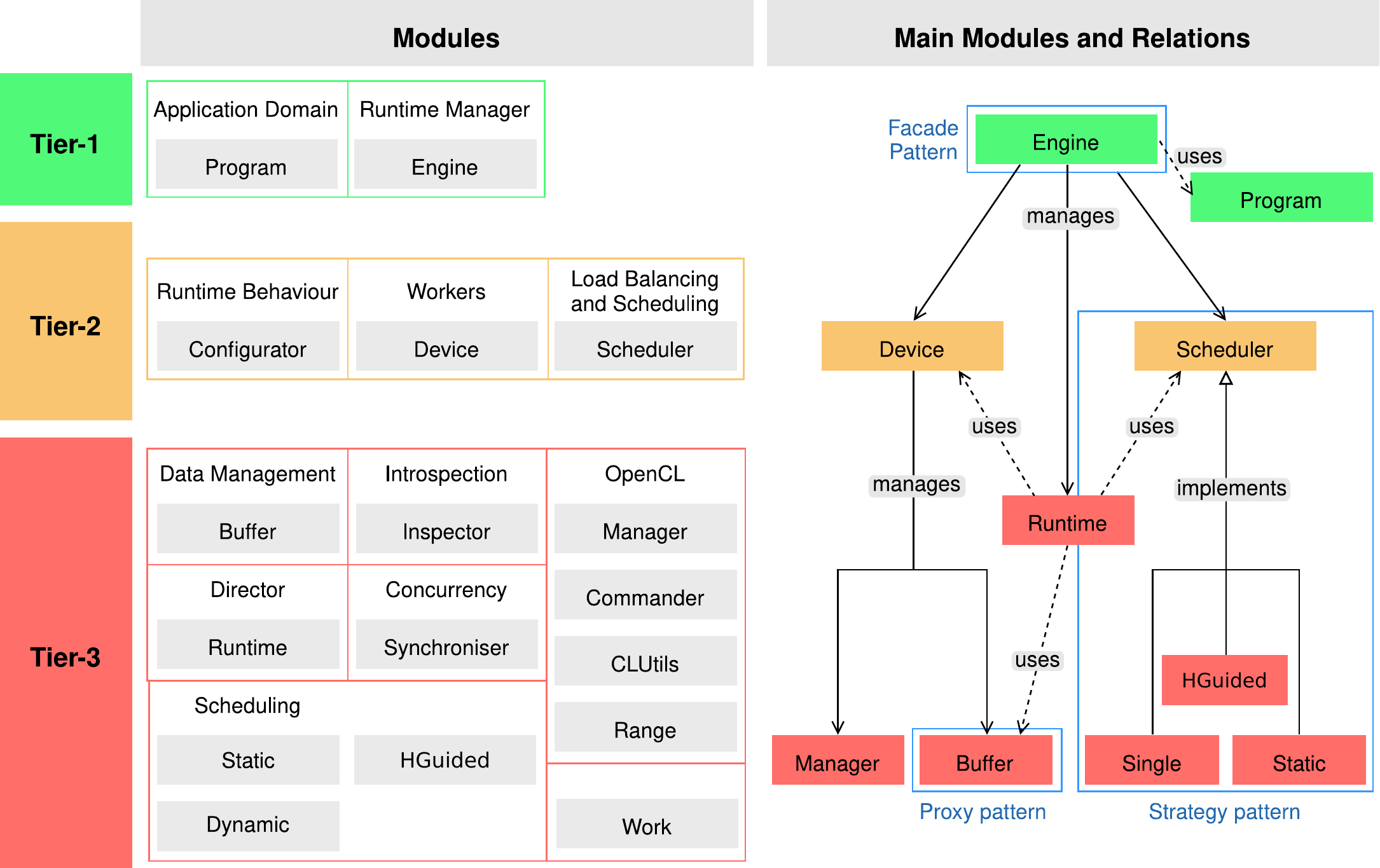}
%   \vspace{-2mm}
   \caption{EngineCL Tiers and modules.}
   \label{fig:tiers}
\end{figure}

%% \vspace{-5mm}
%% \hfill

While OpenCL allows code portability on different types of devices, the programmer is responsible for managing many concepts related to the architecture,
% such as platforms, devices, contexts, buffers, queues, kernels, kernel arguments, data transfers, kernel executions and error control sections.
such as platforms, contexts, buffers, queues or data transfers. % or error control sections.
When the number of devices, operations and data management processes increases, the code grows quickly with OpenCL, decreasing the productivity and increasing the maintainability effort. EngineCL solves these issues by providing a runtime with a higher-level API that efficiently manages all the OpenCL resources of the system. %underlying system.
% independently.
%Despite the increased portability OpenCL offers, the programmers still have to know the architecture of the system in detail and are responsible for the time- consuming task of adapting system management and load balanc- ing to the actual underlying system.

EngineCL redefines the concept of \textit{program} to facilitate its usage and the understanding of a kernel execution. Because a program is associated with the application domain, it has data inputs and outputs, a kernel and an output pattern.
%The data is materialized as \Cpp{} containers (like \textit{vector}), memory regions (C pointers) and kernel arguments (POD-like types, pointers or custom types).
%The kernel accepts directly an OpenCL-kernel string, and the output pattern is the relation between the \textit{global work size} and the size of the output buffer written by the kernel. The default value is $1:1$, because every work-item (thread) writes to a single position in the output buffers ($\frac{1\ out\ index}{1\ work-item}$, e.g. the third work-item writes to the third index of every output buffer).
It is designed to support massive data-parallel kernels, but thanks to the program abstraction the runtime will be able to orchestrate multi-kernel executions, prefetching of data inputs, optimal data transfer distribution, iterative kernels and track kernel dependencies and act accordingly. Therefore, the architecture of the runtime is not constrained to a single program.

%%  As can be seen in Figure \ref{fig:tiers}, the Tier-1 API has been provided mainly because of a Facade Pattern, facilitating the use and readability of the Tier-2 modules, reducing the signature of the higher-level API with the most common usage patterns. The Buffer is implemented as a Proxy Pattern to provide extra management features and a common interface for different type of containers, independently of the nature (C pointers, \Cpp{} containers) and its locality (host or device memory).
%%  %% Currently, it supports host-initialised C pointers and \Cpp{} vector containers, and other types can be easily integrated with this pattern.
%% Finally, the Strategy Pattern has been used in the pluggable scheduling system, where each scheduler is encapsulated as a strategy that can be easily interchangeable within the family of algorithms.
%%  Because of its common interface, new schedulers can be provided to the runtime. % system.

The runtime follows Architectural Principles with well-known Design Patterns to strengthen the flexibility in the face of changes, as can be seen in Figure \ref{fig:tiers}. Tier-1 and Tier-2 are accessible by the programmer. The lower the Tier, the more functionalities and advanced features can be manipulated. Most programs can be implemented in EngineCL with just the Tier-1, by using the \textit{EngineCL} and \textit{Program} classes. The Tier-2 should be accessed if the programmer wants to select a specific \textit{Device} and provide a specialized kernel, use the \textit{Configurator} to obtain statistics and optimize the internal behavior of the runtime or set options for the \textit{Scheduler}. Tier-3 contains the hidden inner parts of the runtime that allows a flexible system regarding memory management, pluggable schedulers, work distribution, high concurrency and OpenCL encapsulation.

\subsection{HGuided}
The architecture allows to easily incorporate a set of schedulers, as it is shown in Figure \ref{fig:tiers}. Three well-known load balancing algorithms, Static, Dynamic and HGuided, are implemented in EngineCL \cite{Nozal:2018} \cite{Maat:2009}. 
Static and Dynamic strategies have their strong points and their weak spots. The HGuided algorithm is an attempt to reduce the synchronization points of the Dynamic while retaining most of its adaptiveness.

HGuided offers a variation over the Dynamic algorithm by establishing how the data-set is divided. The algorithm makes larger packages at the beginning and reduces the size of the subsequent ones as the execution progresses. Thus, the number of synchronization points and the corresponding overhead is reduced, while retaining a small package granularity towards the end of the execution to allow all devices to finish simultaneously.

Since it is an algorithm for heterogeneous systems the size of the packets is also dependent on the computing power of the devices. The size of the package for device $i$ is calculated as follows:

\begin{equation*}
    packet\_size_i = \floor*{\frac{G_r\,P_i}{k_i\,n\,\sum_{j=1}^{n} P_j}}
\end{equation*}
where $k_i$ is an arbitrary constant. The smaller the $k$ constant, the faster decreases the packet size. Tweaking this constant prevents too large packet sizes when there are only a few devices, with cases such as giving half the workload in the first packet to a device, unbalancing the load.
%% Where $k$ is a constant, between 2 and 3, to decrease the package size faster the lower the constant is. This avoids too big package sizes when there are a few devices, with cases like giving half the workload in the first package to a device, imbalancing the load.
$G_r$ is the number of pending work-groups and is updated with every package launch. The parameters of the HGuided are the computing powers and the minimum package size of the devices to be used. The minimum package size is a lower bound for the $packet\_size_i$ and the minimum package sizes are usually dependent on the computing power of the devices, being bigger package sizes in the most powerful devices.

HGuided is optimized by applying a combined tweaking to both the $k$ constant and the minimum package size, inversely related. For each device, a pair of minimum package size and $k_i$ constant is given. The former is a multiplier of the local work size and it increases with more powerful devices, while the latter decreases in such cases. The $k_i$, although related with the computing power of each device, it is established with values between 1 and 4 to avoid crossing the border penalties: neither too large nor too small packages.
%smaller than 1 produces too large packages, greater than 4, too small ones.
% Some generalizations can be applied based on the cases studied, obtaining the best results with 3.5, 1.5 and 1 for the CPU, integrated GPU and discrete GPU, respectively.

\begin{figure}[H]
   \centering
   \includegraphics[width=0.44\textwidth]{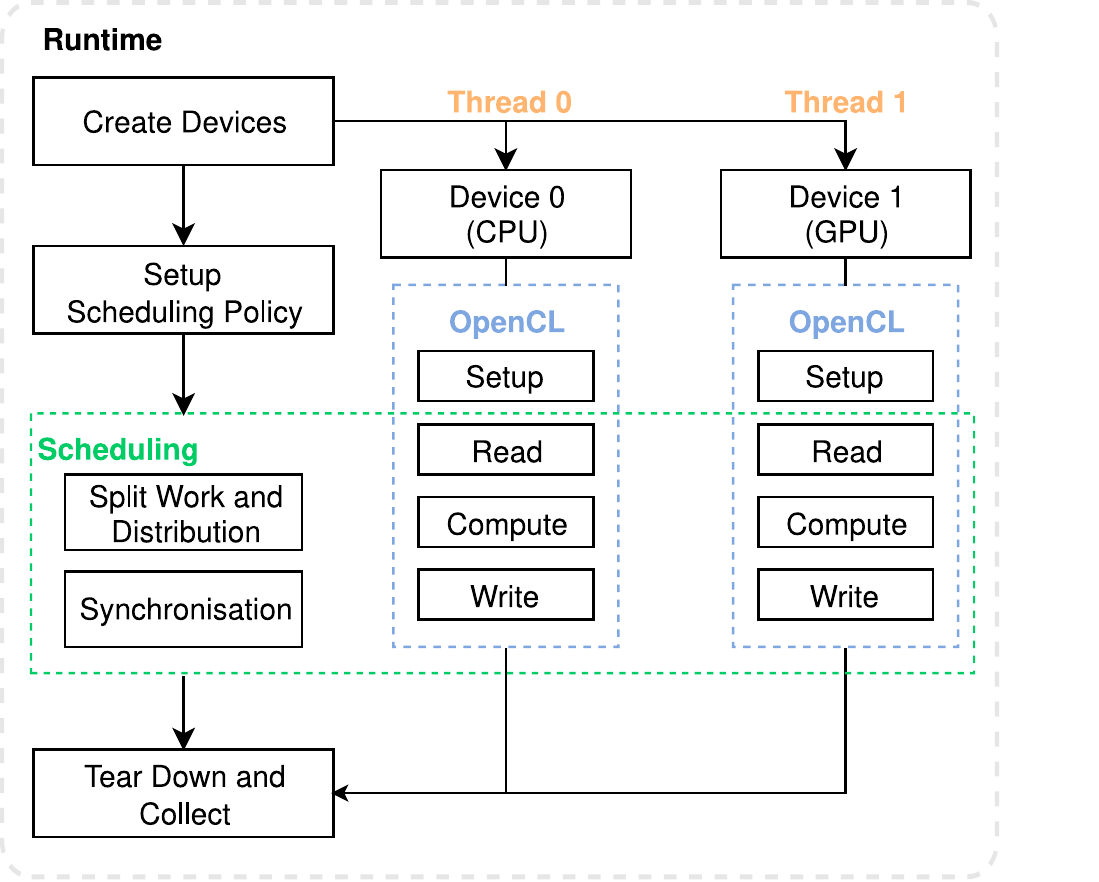}
   \hspace{-5mm}
%   \vspace{10mm}
   \caption{Scheduling overview.}
   \label{fig:overview}
\end{figure}

\section{Optimizations of EngineCL}
\label{sec:Optimization}
EngineCL has been developed in \Cpp{}, mostly using \Cpp{}11 modern features to reduce the overhead and code size introduced by providing a higher abstraction level.
%% Many modern features like \textit{rvalue references}, \textit{initializer lists} and \textit{variadic templates} have been used to provide a better and simpler API, at the same time as efficient management operations are performed inside the runtime.
%% Many modern features such as \textit{rvalue references}, \textit{initializer lists} or \textit{variadic templates} have been used to provide a better and simpler API while preserving efficient management operations internally.
% at the same time as efficient management operations are performed inside the runtime.
When there is a trade-off between internal maintainability of the runtime and a performance penalty seen by profiling, it has been chosen an implementation with the minimal overhead in performance.

%JLB: LA descomposición en Tiers, ya se ha comentado un poco más arriba. Se puede decir: "Como s eha explicado en la sección anterior...". Además hay que volver a referenciar la figura 2, ya que hemos cambiado de sección.
% NOZAL: no sé si ambas son de diseño.

%% The implementation follows feature-driven development to allow incremental features based on requested needs when integrating new vendors, devices, type of devices and benchmarks.

%%JLB: Lo quito para WEHA
%Implementation techniques are profiled with a variety of OpenCL drivers from the major vendors and versions, but also in devices of different nature, such as integrated and discrete GPUs, CPUs and accelerators \cite{jos:2019}.

EngineCL has a multi-threaded architecture that combines the best measured techniques regarding OpenCL management of queues, devices and buffers. Some of the decisions involve atomic queues, paralleled operations, custom buffer implementations, reusability of costly OpenCL functions, efficient asynchronous enqueueing of operations based on callbacks and event chaining. These mechanisms are used internally by the runtime and hidden from the programmer to achieve efficient executions and transparent management of devices and data.

Figure \ref{fig:overview} depicts a simplified view of the scheduling and work distribution, while showing a key design in the architecture: the low-level OpenCL API is encapsulated within the concept of Device, managed by a thread.

Two design decisions have been affected to support optimizations that reduce overheads produced both in the initialization and closing stages of the program, mainly due to the use of OpenCL drivers for commodity infrastructures. These modifications are tagged as \textit{initialization} and \textit{buffer optimizations}.

The first optimization focuses on taking advantage of the discovery, listing and initialization of platforms and devices by the same thread (Runtime). In parallel, both the thread in charge of load balancing (Scheduler) and the threads associated with devices (Device) take advantage of this time interval to start configuring and preparing their resources as part of the execution environment. These threads will wait only if they have finished their tasks independent of the OpenCL primitives, instantiated by the Runtime. The runtime takes advantage of the same discovery and initialization structures to configure the devices before delegating them to the Scheduler and Device threads, which will be able to continue with the following stages. These optimizations reduce the execution time affecting the beginning and end of the program, due to the increase of the parallel fraction of the program as well as the reuse of the structures in memory, liberating the redundant OpenCL primitives.

On the other hand, some modifications have been made when instantiating and using both input and output buffers (Buffer). The variety of architectures as well as the importance of sharing memory strategies save costs when doing transfers and unnecessary complete bulk copies of memory regions, usually between main memory and device memory, but also between reserved parts of the same main memory (CPU - integrated GPU). By tweaking OpenCL buffer flags that set the direction and use of the memory block with respect to the device and program, OpenCL drivers are able, if possible, to apply underlying optimizations to the memory management.

%% TODO(future)
%% \begin{figure*}[!htb]
%%    \centering
%%    \includegraphics[width=120mm]{assets/enginecl_api.png}
%%    \caption{EngineCL API.}
%%    \label{fig:translation}
%% \end{figure*}
%% \vspace{-2mm}

%% file: sections/methodology.tex
\newcolumntype{M}{>{\begin{varwidth}{18mm}}l<{\end{varwidth}}} %M is for Maximal column
\section{Methodology}
\label{sec:Methodology}

The experiments are carried in a machine composed of an AMD A10-7850K APU and Nvidia GeForce GTX 950 GPU. The CPU has 2 cores and 2 threads per core at 3142 Mhz with only two cache levels, exposing 4 OpenCL compute units. The APU's on-chip GPU is a GCN 2.0 Kaveri R7 DDR3 with 512 cores at 720 Mhz with 8 compute units, denoted as {\em iGPU} from now on. Finally, the Nvidia discrete GPU has 768 cores at 1240 Mhz with GDDR5, providing 6 compute units, denoted as {\em GPU}.

Five benchmarks have been used to show a variety of computations regarding the performance gains when multiple heterogeneous devices are co-executed. Table \ref{tbl:properties} shows the properties of every benchmark. Gaussian, Binomial, Mandelbrot and NBody are part of the AMD APP SDK, while Ray, providing two different scenes, is an open source implementation of a Raytracer \cite{EngineCLBenchsuite}. They provide enough variety in terms of OpenCL development issues, regarding many parameter types, local and global memory usage, custom struts and types, number of buffers and arguments, different local work sizes and output patterns.

\begin{table}[h]
\begin{center}
\caption{Benchmarks and their variety of properties.}
\label{tbl:properties}
%% \vspace{-2mm}
  %% \footnotesize
\scriptsize
\renewcommand{\arraystretch}{2}
%% \ssmall
%%\setlength\extrarowheight{2pt}
\begin{tabular}{@{\hskip 1mm}M@{\hskip 2mm}@{\hskip 2mm}>{\columncolor[gray]{0.95}}c@{\hskip 2mm}@{\hskip 2mm}c@{\hskip 2mm}@{\hskip 2mm}>{\columncolor[gray]{0.95}}c@{\hskip 2mm}@{\hskip 2mm}c@{\hskip 2mm}@{\hskip 2mm}>{\columncolor[gray]{0.95}}c@{\hskip 2mm}}%p{18mm}|p{18mm}|p{20mm}|p{25mm}|p{15mm}|}
  %% \rowcolor{light-gray}
  %% {\textbf{Property}}
  \parbox[c][1.7cm]{2.2cm}{\vspace{1.2cm}\textbf{Property}}
  & \rotatebox[origin=s]{65}{\centering\textbf{Gaussian}}
  & \rotatebox[origin=c]{65}{\centering\textbf{Binomial}}
  & \rotatebox[origin=c]{65}{\parbox[c]{13mm}{\centering\textbf{NBody}}}
  & \rotatebox[origin=c]{65}{\parbox[s]{13mm}{\centering\textbf{Ray}}}
  & \parbox[c][1.8cm]{0.7cm}{\rotatebox[origin=c]{65}{\centering\textbf{Mandelbrot}\vspace{3mm}}}
  \\ \hline
            Local Work Size    & 128    & 255     & 64     & 128    & 256     \\[1pt] \hline
            Read:Write buffers & 2:1    & 1:1     & 2:2    & 1:1    & 0:1     \\[1pt] \hline
            Out pattern        & 1:1    & 1:255   & 1:1    & 1:1    & 4:1     \\[1pt] \hline
            Kernel args        & 6      & 5       & 7      & 11     & 8       \\[1pt] \hline
            Use local memory   & no     & yes     & no     & yes    & no      \\[1pt] \hline
            Use custom types   & no     & no      & no     & yes    & no      \\[1pt] \hline
            Size               & 8192px & 4194304 & 229376 & 4096px & 14336px \\[1pt] \hline
            Other params       & 31px   &         &        & scene  & 5000    \\[1pt] \hline

%% other arguments & filter size 91x91 & scene with 10 objects and 3 lights & 255 steps & 524288 iterations & \\ \hline
%% problem size & image width and height & image width and height & options & image width and height & num bodies\\ \hline
\end{tabular}
\end{center}
\end{table}

% TODO faltaría explicar el estudio sobre tamaños de problema (si nos interesa)

The performance evaluation of EngineCL is done by analyzing the co-execution when using different scheduling configurations in a heterogeneous system composed of three different devices.
Each program uses a single problem size, given by the completion time of around 2 seconds in the fastest device (GPU).

The scheduling configurations are grouped by algorithm. The first two bars represent the Static algorithm varying the order of delivering the packages to the devices. The one labelled \emph{Static} delivers the first chunk to the CPU, the second to the iGPU and the last one to the GPU, while in the \emph{Static rev} the order is \textit{GPU-iGPU-CPU}. The next three show the Dynamic scheduler configured to run with 64, 128 and 512 chunks. Finally, the latter two present the HGuided algorithm and its new optimized version.
% Traer de resultados las explicaciones de static, static rev
% Mencionar las configuraciones de planiificación

To guarantee integrity of the results, 50 executions are performed per case. An initial execution is discarded for every set of iterations to avoid warm-up penalties in some OpenCL drivers and devices.

%The problem sizes changed for each device are the image size for Gaussian, Ray and Mandelbrot, the number of options for Binomial and the the number of bodies for NBody.

To evaluate the performance of the load balancing algorithms the total response time is measured, including kernel computing and buffer operations (reading and writing), but excluding program initialization and releasing.

Three metrics are used to evaluate the performance of EngineCL: balance, speedup and efficiency.

To measure the effectiveness of load balancing, we calculate the balance as $ \frac{ {T_{FD}} }{ {T_{LD}} }$, where $T_{FD}$  and $T_{LD}$ are the execution time of the device that finish at first and last, respectively. Thus, it is $1$ if all finish at the same time.

For the latter two metrics, the baseline is always the fastest device running a single invocation of the kernel (GPU). 
%The speedup is calculated as the ratio between the execution time on the baseline and on the heterogeneous system.
Due to the heterogeneity of the system and the different behaviour of the benchmarks, the maximum achievable speedups depend on each program. These values are derived from the response time $T_i$ of each device:
% TODO meter la n aquí
\vspace{-2mm}
\begin{equation*}
   S_{max} =  \frac{1}{max_{i=1}^n\{T_i\}} \sum_{i=1}^n T_i
   \label{smax}
\end{equation*}

Additionally, the efficiency of the heterogeneous system has been computed as the ratio between the maximum achievable speedup and the empirically obtained speedup for each benchmark. $Eff = \frac{S_{real}}{S_{max}}$.

%% file: sections/results.tex
\begin{figure*}[t] % !b !t
\centering
\includegraphics[width=0.49\textwidth,trim={5mm 5mm 6mm 5mm},clip]{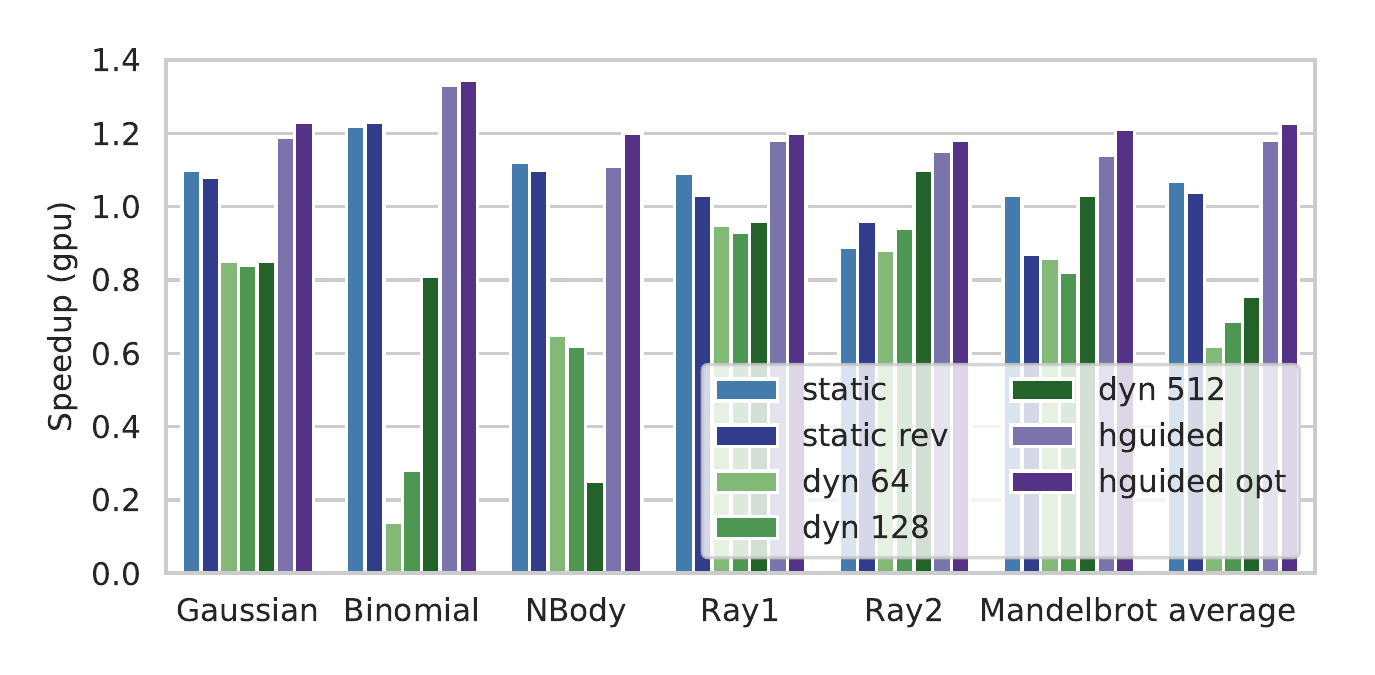}
%\frame{
\hfill
\includegraphics[width=0.49\textwidth,trim={5mm 5mm 6mm 5mm},clip]{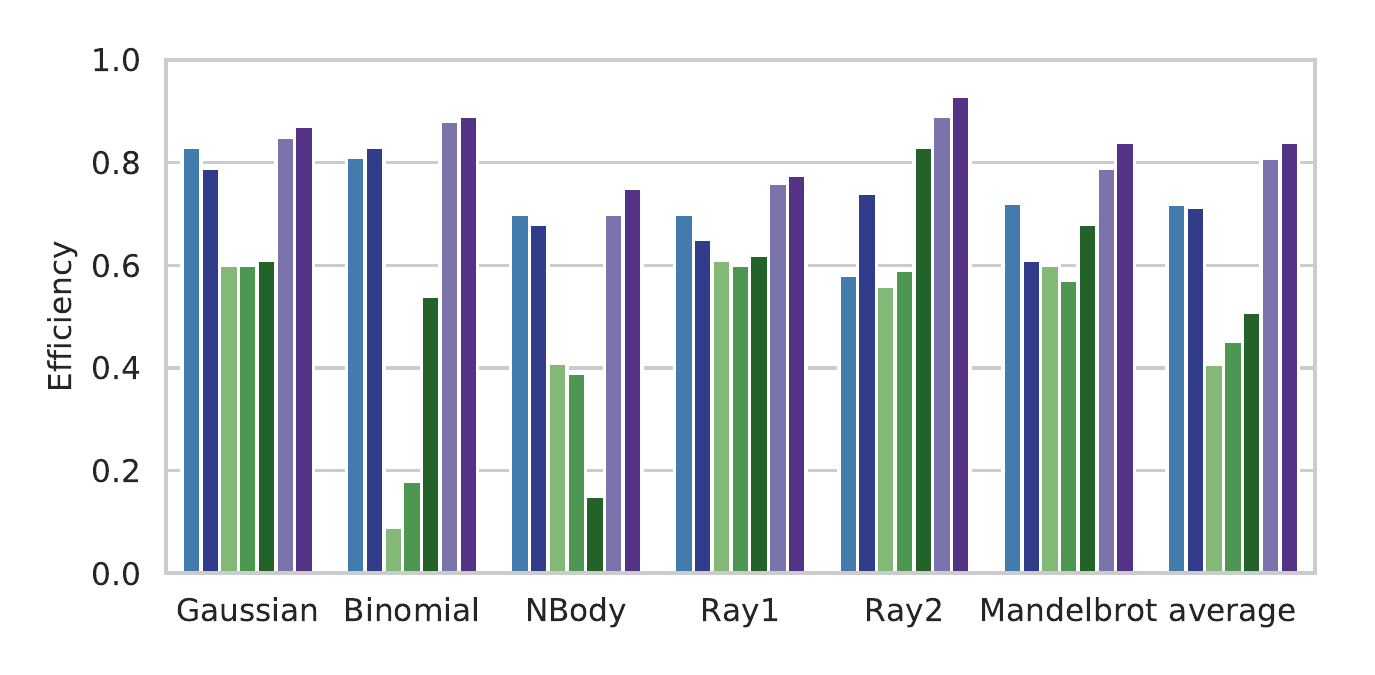}
%}
%\includegraphics[width=0.32\textwidth,trim={7mm 7mm 16mm 10mm},clip]{charts/summary/balance.pdf}
%\caption{Speedups (left), Efficiency (center) and Balance (right) for every scheduler compared with a single GPU.}
  \vspace{-2mm}
  \caption{Speedups (left) and Efficiency (right) for every scheduler and program compared with a single GPU.}
  \label{fig:speedupEfficiencyApulab}
\end{figure*}

\section{Experimental Results}
\label{sec:ExperimentalResults}
This section presents results of experiments carried out to evaluate the performance when using all the devices in the system with the load balancing algorithms, including the new runtime and HGuided optimizations.

\subsection{Performance Results}

The performance results achieved in the heterogeneous system with different load balancing algorithms are shown in Figure \ref{fig:speedupEfficiencyApulab}, where the speedups and efficiency are depicted. 
%The ordinate axis shows the speedup and efficiency of the heterogeneous system for every scheduling configuration and benchmark.
The abscissa axis contains the benchmarks defined in Section \ref{sec:Methodology}, each one with seven scheduling configurations. The last group of bars shows the average (geometric mean) per scheduler. 
% TODO: reescribir idnicando la GPU a 1.0 y por debajo es mejor no coejecución
%JLB: These variations are presented to show the behaviours of the algorithms in different scenarios. 

%JLB: Todo esto se dice en la metodología. Como andamos mal de espacio lo quito. 
%The speedups presented in this section are due to the co-execution of the benchmarks on the CPU, iGPU and GPU simultaneously, compared with only using the fastest device, the GPU. The efficiency gives an idea of how well a load is balanced. A value of 1.0 represents that all the devices have been working all the time, thus achieving the maximum obtainable speedup.
The figures reveal that, for all benchmarks, HGuided achieves the best results. Classifying the programs as regular (Gaussian, Binomial and Nbody) and irregular (Ray and Mandelbrot), there is a tendency that shows how the Static is better for the former, while the Dynamic for the latter. In any case, the Static would be the second best option, indicating the importance of taking into account the computing power of each device in the heterogeneous system, feature not present in the dynamic algorithm. The results show how the dynamic is penalized when an inappropriate number of packages is selected, as can be seen with bigger package sizes in Binomial, Ray2 and Mandelbrot, and smaller ones in NBody.

These overheads introduced by Dynamic are mostly mitigated when using HGuided, being always better than using the fastest device (GPU), with average efficiencies of 0.81 and 0.84, when using the default and the new optimized version, respectively. The previous version of HGuided obtained the best speedups on average compared with Static and Dynamic, except in NBody, where a Static combination obtains the same performance. Due to the optimizations applied, the results of the HGuided improve 3\% for regular problems and 3.5\% for the irregular ones, being ahead of the rest scheduling configurations and reaching efficiencies of up to 0.89 and 0.93 for Binomial and Ray2, respectively.

\noindent\begin{minipage}{.49\textwidth}
             \begin{figure}[H] % !t
                 \centering
                 \includegraphics[width=0.99\textwidth,trim={4mm 4mm 6mm 4mm},clip]{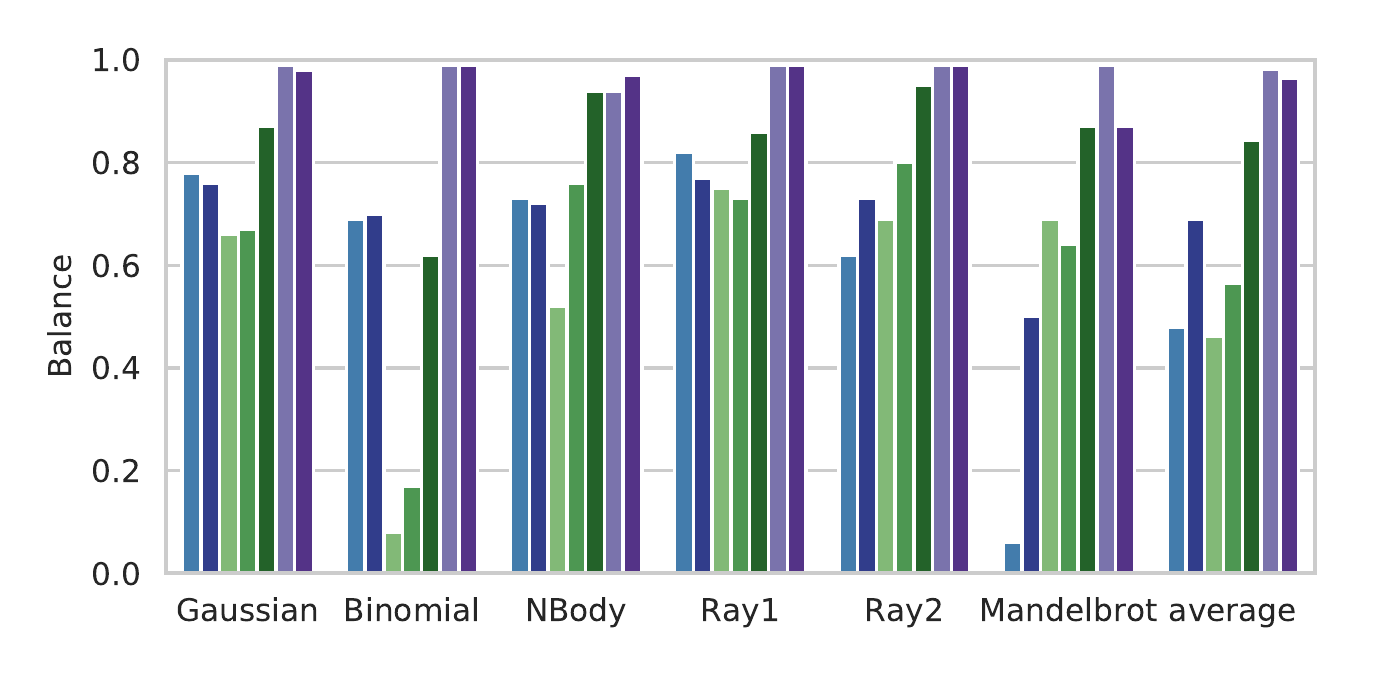}
                 \vspace{-5mm}
                 \hyphenation{compared}
                 \caption{Balance for every scheduler compared with a single GPU.}
                 \label{fig:balanceApulab}
                 \vspace{3mm}
             \end{figure}
\end{minipage}

Figure \ref{fig:balanceApulab} depicts the balance metric obtained per scheduler, as it is defined in Section \ref{sec:Methodology}, achieving near the best balance when using HGuided, in all applications. This is a consequence of the computation of smaller packages at the end of the execution. For regular problems, the balance is directly related with the performance of its scheduling configuration, but it is not completely fulfilled for irregular ones, with cases like Mandelbrot. It has higher performance in Static than in its reversed version, but it suffers imbalance. These variations are produced because a slow device (CPU) finishes working and no more packages need to be computed, waiting for the other devices, but finishing the total problem computation in less time, because faster devices compute more work.

\begin{figure*}[t] % !t
    \centering
    \includegraphics[width=0.30\textwidth,trim={7mm 3mm 0mm 6mm},clip]{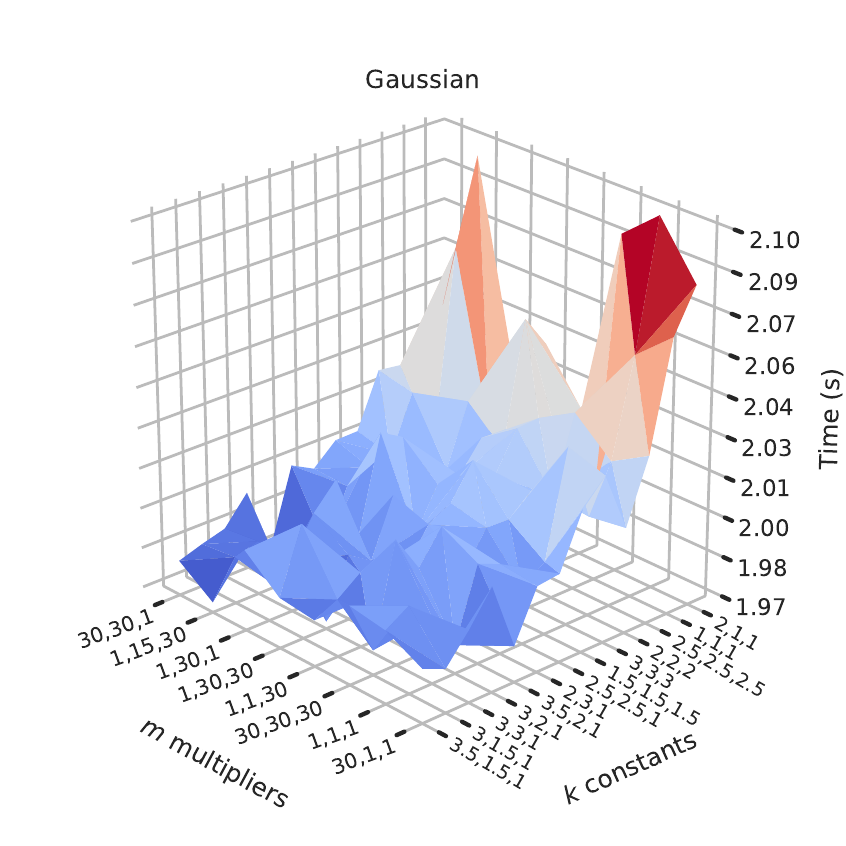}
    \includegraphics[width=0.30\textwidth,trim={7mm 3mm 0mm 6mm},clip]{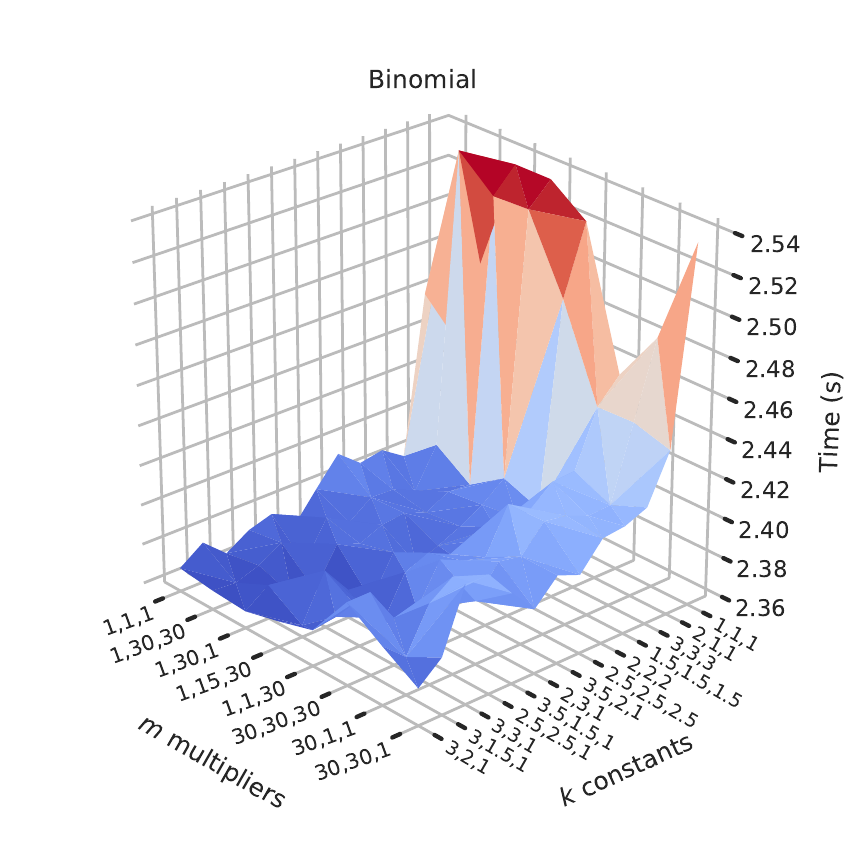}
    \includegraphics[width=0.30\textwidth,trim={7mm 3mm 0mm 6mm},clip]{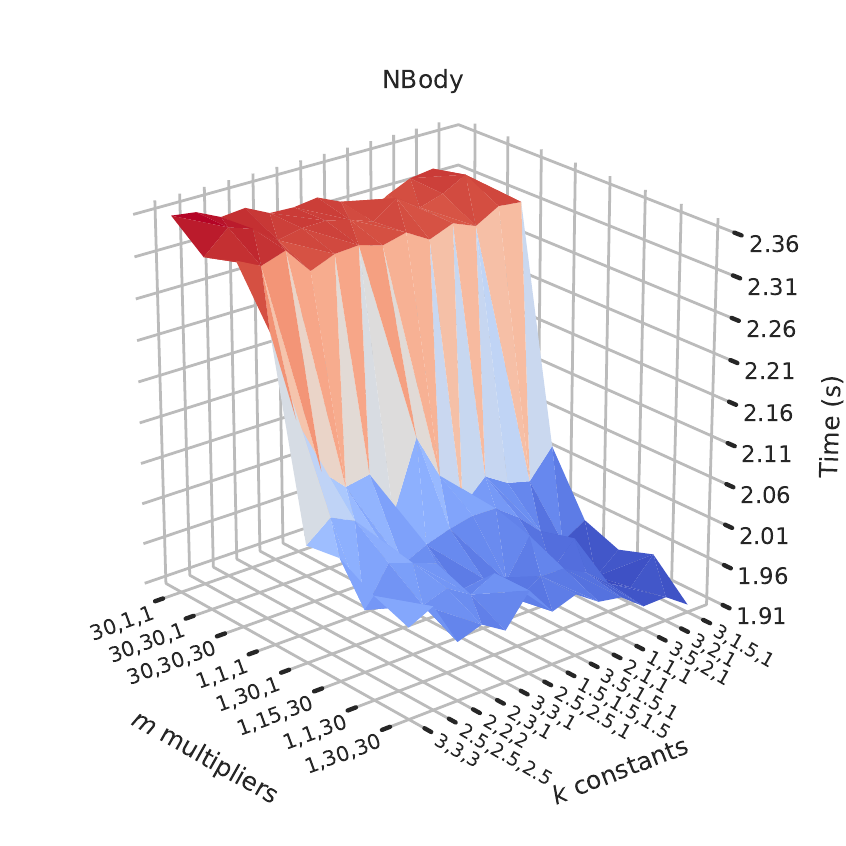}
    \includegraphics[width=0.30\textwidth,trim={7mm 3mm 0mm 6mm},clip]{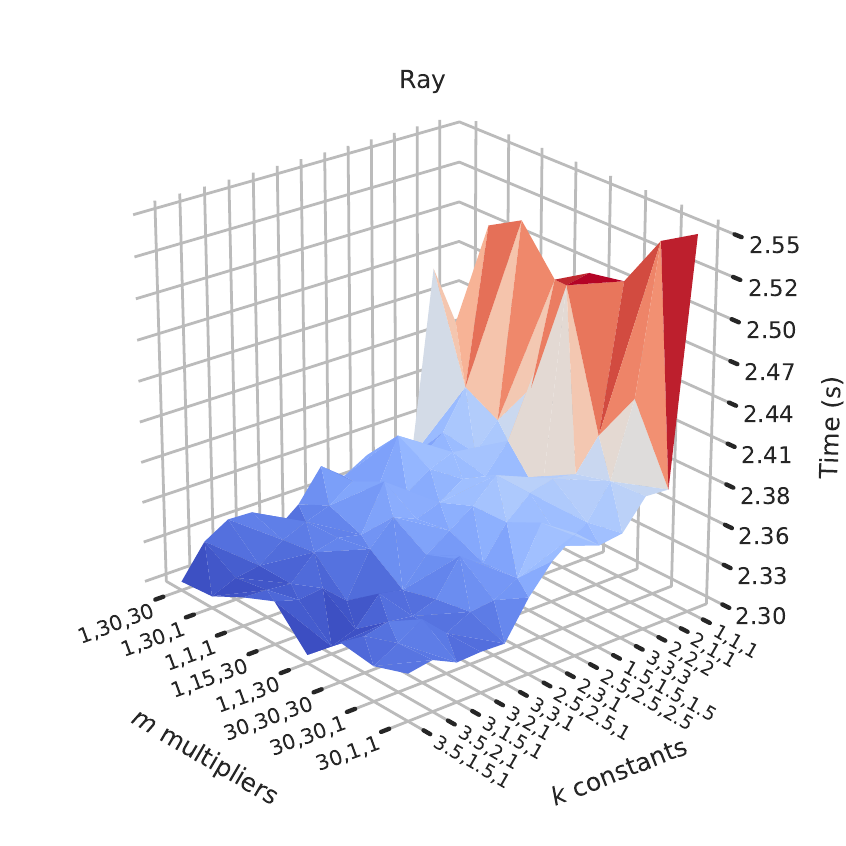}
    \includegraphics[width=0.30\textwidth,trim={7mm 3mm 0mm 6mm},clip]{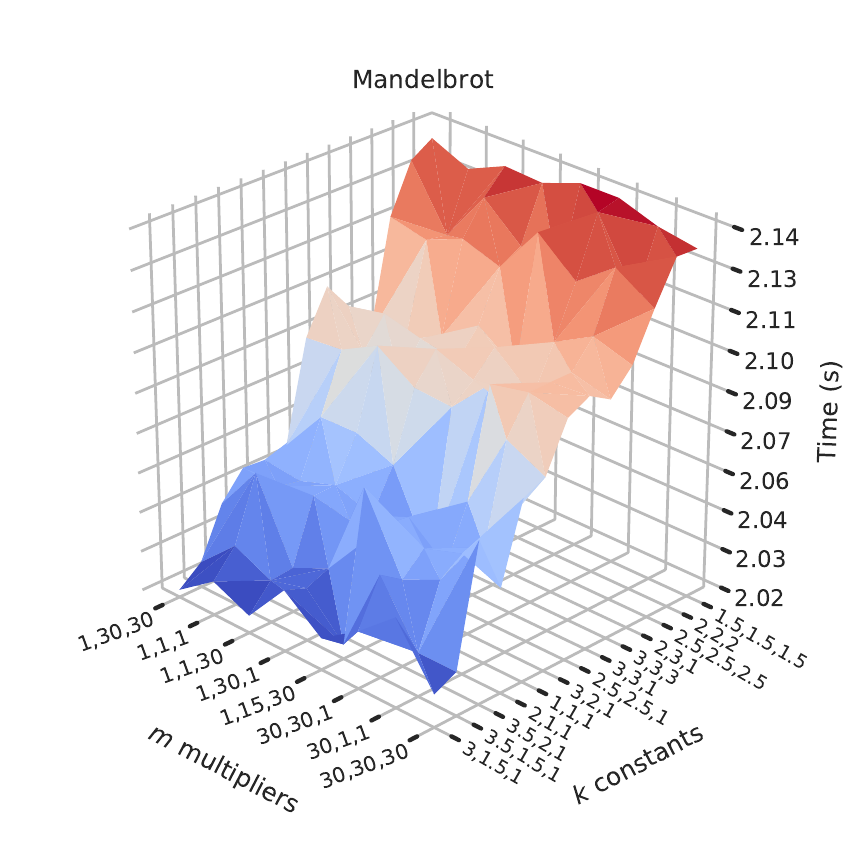}
%    \vspace{-2mm}
%    \hyphenation{compared}
    \caption{HGuided scheduler performance: $m$ multiplier (minimum package size) and $k$ constant parameter combination.}
    \label{fig:matrixHGuidedParams}
    \vspace{-3mm}
\end{figure*}

\subsection{Optimizations Evaluation}

Taking into consideration the global vision of the problem as well as the excellent results obtained with the new proposal of the HGuided, it is necessary to emphasize the optimization work performed at the algorithmic level as well as in the runtime, as it was detailed in Section \ref{sec:Optimization}.

The changes in the HGuided algorithm introduce enough variability to obtain an explosion of combinations in the search for the most optimal parameters, both in average and maximum cases per problem. Figure \ref{fig:matrixHGuidedParams} shows how the performance is affected by the combination of pairs $m, k$, multiplier value to obtain the minimum package size and $k$ constant, both for each device in the heterogeneous system.
The Z axis sets the program execution time for the problem size indicated in the Section \ref{sec:Methodology}. The X and Y axes show the multiplier for minimum package size and $k$ constant for each device, respectively. The order of the three values represent the CPU, iGPU and GPU. For example, a Gaussian execution with the values $k 3,3,1$ and $m 1,15,30$ means that the HGuided has scheduled the workload by distributing smaller packages to the CPU and iGPU but larger to the GPU. This $k$ variation affects the first packages delivered. In addition, as packages are distributed and their size decrease is when the minimum package size limitation occurs. In this case, because of the $m$ selected values, the CPU is not limited, but the iGPU and GPU minimum package sizes are 15 and 30 times larger than the lws (128 in Gaussian), respectively.

For all the programs there is a correlation between the $m$ multipliers and $k$ constants used, inversely related. The charts show how the $k$ constants have a greater impact on performance than the minimum packet size, except in NBody, where the limitation of the minimum package size for the CPU is completely relevant to avoid synchronization overheads. The more packages are created, the more management needs to be performed, and both Runtime and Scheduler units are CPU-managed (host thread), incurring in more overheads when dealing with transfers and computations. Although this effect can be appreciated in every application, in NBody it is even more highlighted due to its communications.

Considering all programs, by classifying the results based on the performance average, several conclusions can be extracted to improve the general efficiency: a) the more powerful the device, the greater the minimum package size; b) the more powerful the device, the fewer the $k$ constant; c) there is not a perfect choice for every program, but the combination of $m_i = \{1, 15, 30\}$ and $k_i = \{3.5, 1.5, 1\}$ give the best results; d) if a single $k$ constant should be selected for every device, $k = 2$ is the best option; e) if the CPU is involved in the computation and no previous profiling can be performed, it should maintain $m = 1$ to avoid major penalties in performance.

On the other hand, taking into account the runtime, two tasks have been carried out: the \textit{initialization} optimization affects the execution of the complete program (binary), while the \textit{buffers} optimization improves both binary and based on the region of interest (ROI). The ROI discards the initialization and release stages, considering only the transfer and computation, where a minor management overhead or synchronization call highly penalizes the general performance.

%typically found in bigger and more complex programs,  programs is typically performed in  (ROI, without initialization and release).

\begin{figure*}[t] % !t
    \centering
    \includegraphics[width=0.30\textwidth,trim={2mm 2 2 2},clip]{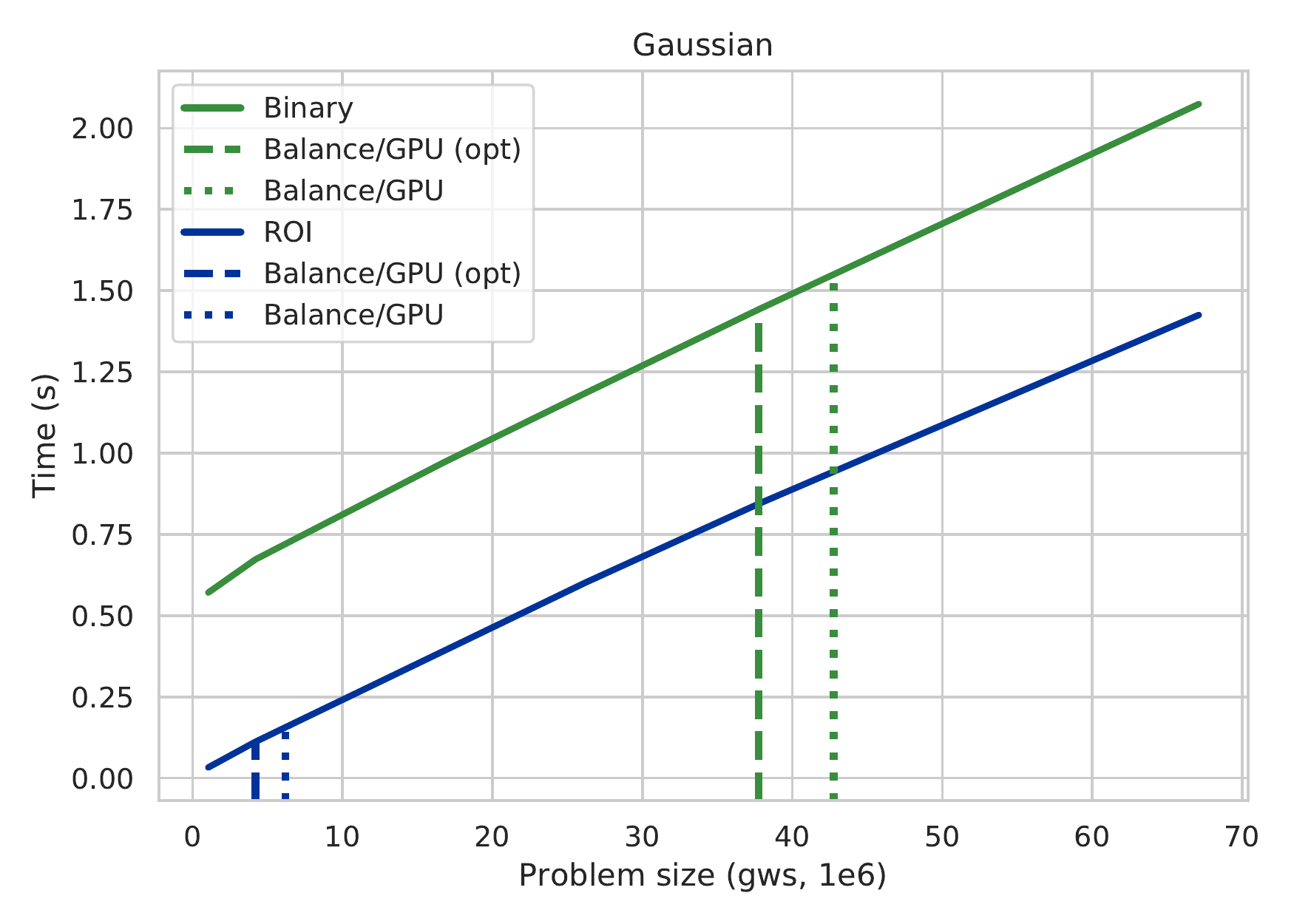}
    \includegraphics[width=0.30\textwidth,trim={2mm 2 2 2},clip]{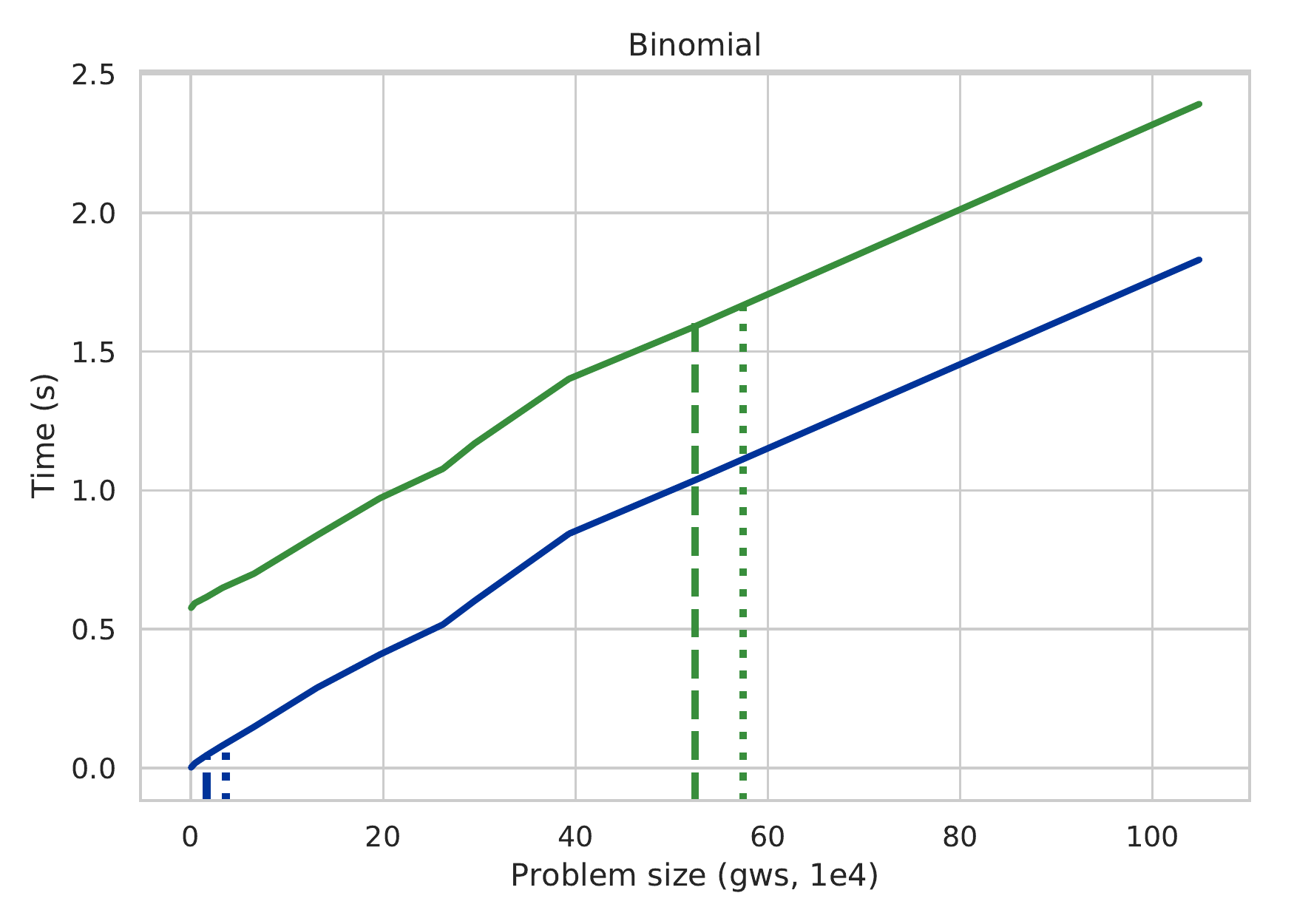}
    \includegraphics[width=0.30\textwidth,trim={2mm 2 2 2},clip]{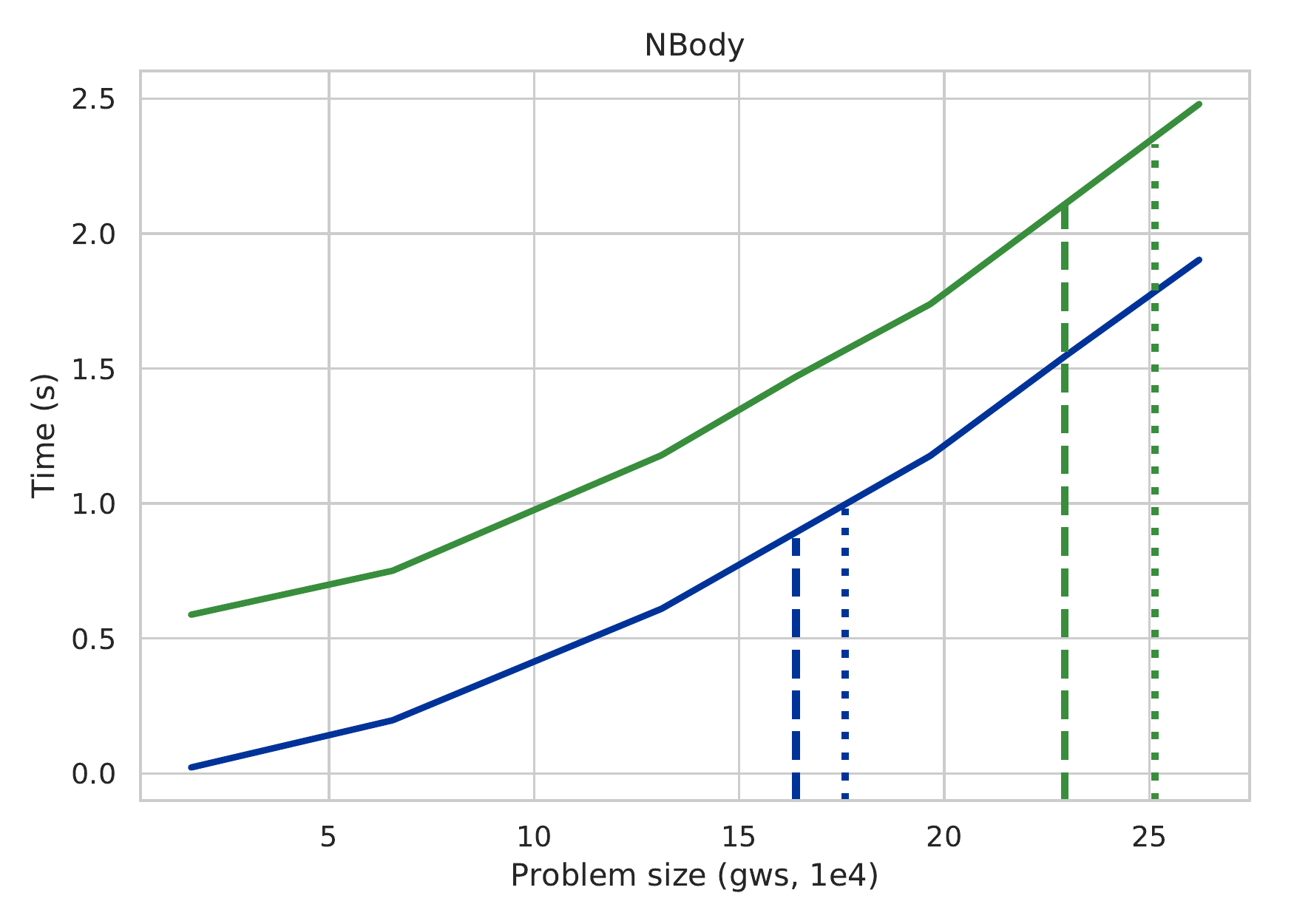}
    \includegraphics[width=0.30\textwidth,trim={2mm 2 2 2},clip]{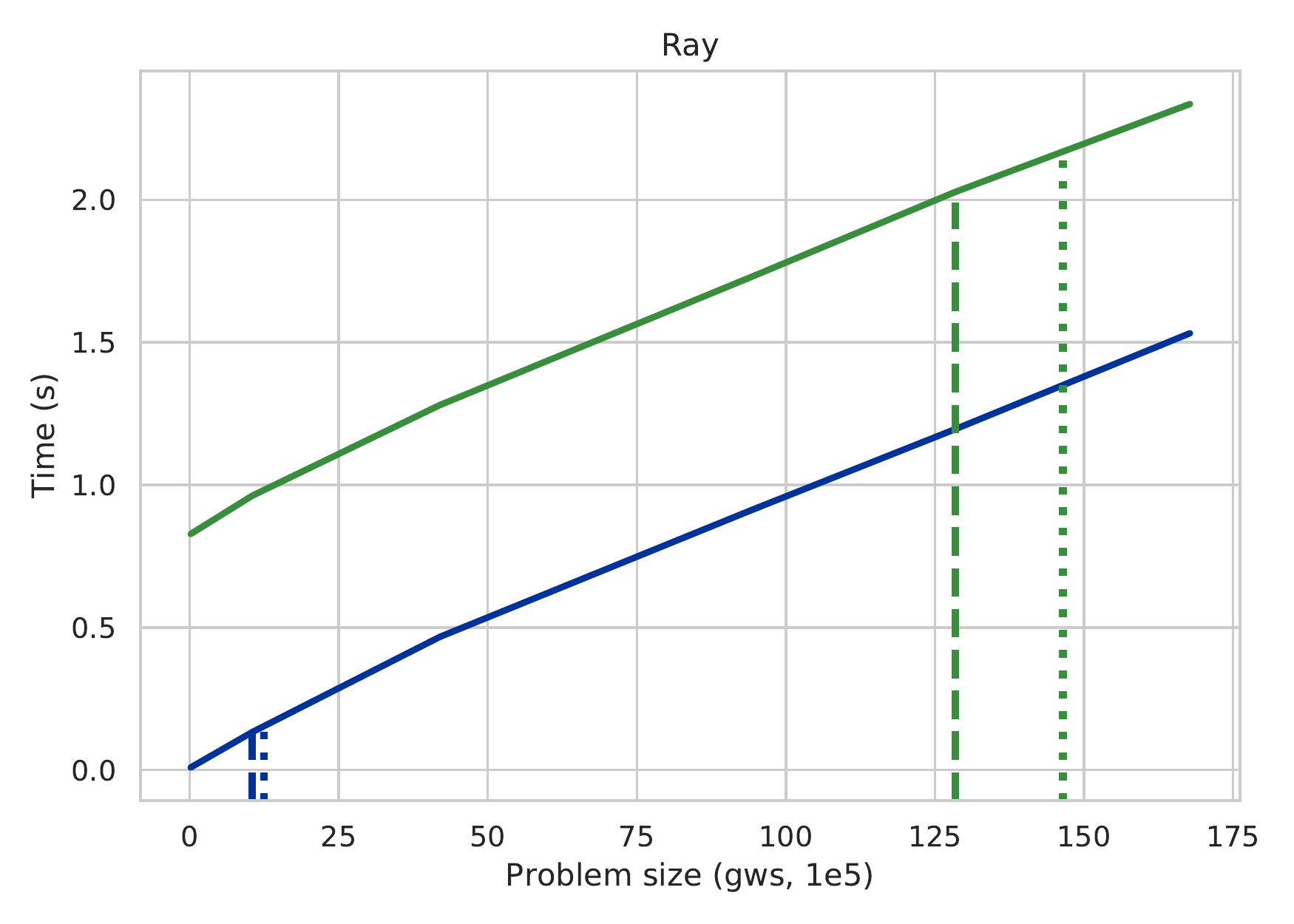}
    \includegraphics[width=0.30\textwidth,trim={2mm 2 2 2},clip]{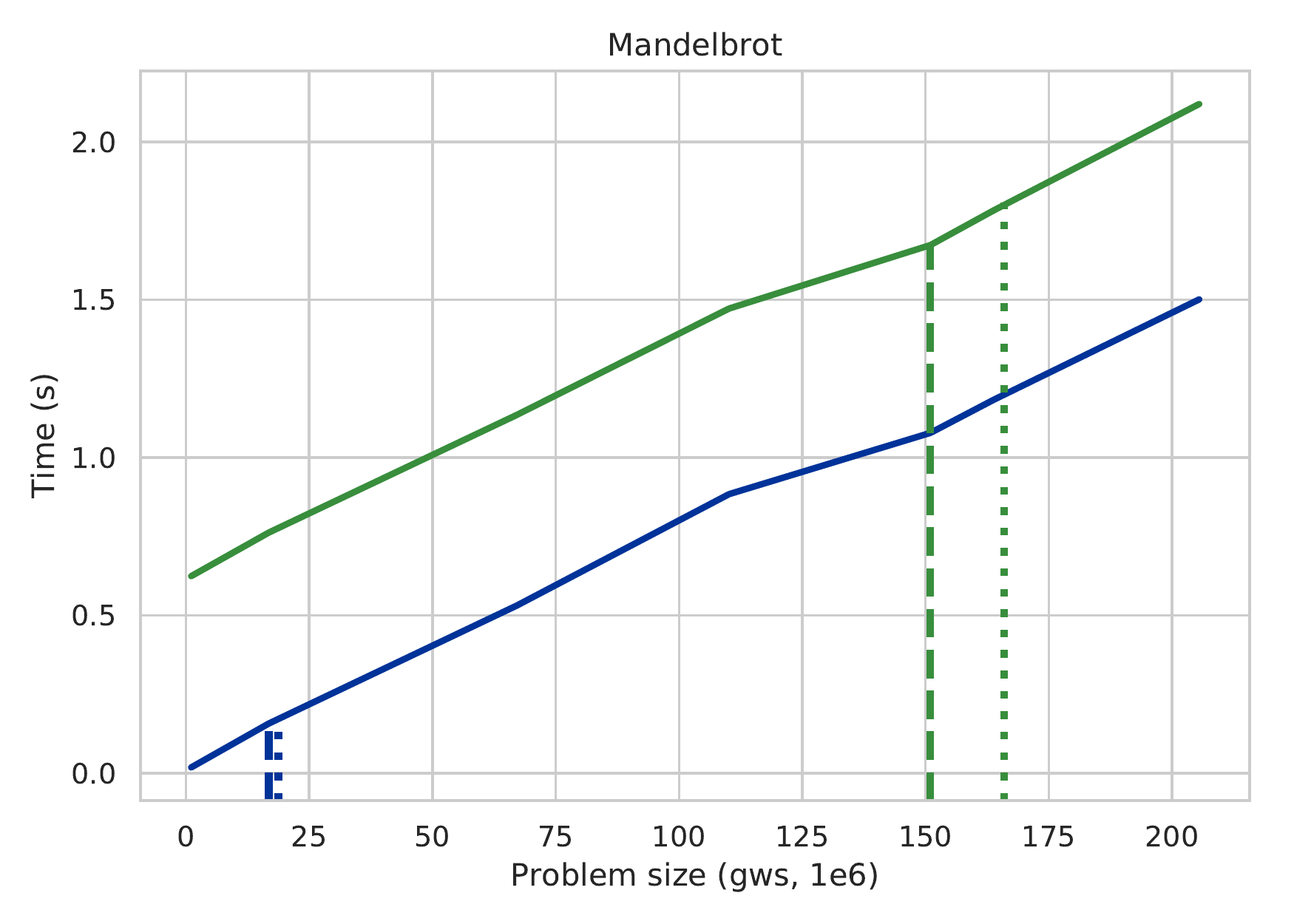}
    %    \vspace{-2mm}
    %    \hyphenation{compared}
    \caption{HGuided execution time per problem size (global work size units) when launching the binary and only the region of interest (transfer and compute). Vertical lines cross the inflection point when load balancing is better than executing in the fastest device (GPU) and how the runtime optimizations affect this trade-off.}
    \label{fig:runtimeOpts}
\end{figure*}

Figure \ref{fig:runtimeOpts} depicts the evolution of the execution time as the problem size increases for each program, as well as the trade-off between single or multi-device execution. The abscissa axis represents the problem size expressed as total work items (gws). The green (top) and blue (bottom) lines show how the execution time increases as the size of the problem grows, for binary and ROI, respectively. Every vertical line represents a inflection point when it crosses the previous continuous lines associated with its execution type, indicating when it is worthwhile to balance the load with HGuided than execute only in the fastest device (GPU). The dotted lines show the previous version, while the dashed lines the new optimized one.

Analyzing the results in detail, all programs show a similar increasing linear trend in their first seconds of execution, except NBody, which grows exponentially. The difference between executions is practically a constant value, due to the amount of common operations during initialization and end of execution, although influenced by the type of kernel (compilation and arguments configuration) but also the amount and size of buffers involved.

The \textit{initialization} optimization has a strong impact on its constant, saving 131 milliseconds on average when reusing OpenCL primitives and the configuration stages are highly parallelized until they are limited by their own dependencies (Buffer, Context, Queue, etc). On the other hand, \textit{buffers} optimization affects both execution modes, but has a greater impact on ROI, as it affects transfers. Devices that share the same main memory can reuse the buffers without penalty. Nevertheless, its impact is minimal for such small problem sizes, since the transfers are practically negligible compared to the rest operations.

The inflection points improve, on average and taking into account the two types of execution, 7.5\% when optimizing the initialization and 17.4\% when facilitating the recognition of buffers types and avoiding unnecessary copies. These two optimizations are fundamental considering the scenario in which these applications are evaluated.

Considering the experimental setup, a set of conclusions can be indicated: a) the average time it is worthwhile balancing the load has to exceed 15 milliseconds when considering the region of interest, and 1.75 seconds when executing the complete program; b) the bigger the problem size, the better it performs load balancing, and therefore, it excels over a single execution over the fastest device; c) the amount of potential optimizations and time savings that can be achieved in the runtime are limited by the gains obtained by applying optimizations on the slowest device and being executed alone (single mode).

In summary, we can conclude that both EngineCL and its load balancing algorithms, thanks to the exhaustive analysis and evaluation, support optimizations and fine-grained work focused on specific scenarios, significantly improving their results.

%% file: sections/related-work.tex
\section{Related Work}
\label{sec:Related}

%JLB: TODO: Se puede dar un poco más de detalle sobre los artículos comentados. Buscar algún trabajo adicional.
%NOZAL: TODO: lo podré mejorar algo
Similar efforts have been performed to integrate a variety of devices to compute a single data-parallel kernel, but the optimizations are focused at the device level by using more complex and problem-specific architectures \cite{Guzman:2018,Guzman:2019}.

Regarding the trade-off between usability and performance, there are projects aiming at high-level parallel programming in \Cpp{}, but most of them provide an
% \Cpp{}
API similar to the STL to ease the parallel programming, like Boost.Compute \cite{Boost.Compute:2016}, HPX \cite{HPX:2017}, Thrust \cite{Thrust:2009}, SYCL \cite{SYCL:2017} and the \Cpp{} Extensions for Parallelism TS \cite{ParallelSTL:2015}.
Thrust, tied to CUDA devices, HPX, which extends C++ Parallelism TS with
% with asynchronous algorithm and additional
future types for distributed computing, or \Cpp{} TS are not OpenCL-centered.
%% While Thurst is tied to CUDA devices, HPX and the \Cpp{} Technical Specification are not focused on OpenCL, but projects like HPX.Compute \cite{HPX.Compute:2017} and SYCLParallelSTL \cite{SyclParallelSTL:2015} provide backends for OpenCL via SYCL.

Projects like HPX.Compute \cite{HPX.Compute:2017} and SYCLParallelSTL \cite{SyclParallelSTL:2015} provide backends for OpenCL via SYCL. % SYCLParallelSTL uses SYCL to showcase ParallelSTL on CPU and GPU.
SYCLParallelSTL exposes ParallelSTL on CPU and GPU.
%, while \Cpp{} TS only targets CPUs.
Proposals like SkelCL \cite{SkelCL:2011} and SkePU \cite{SkePU:2010}
provide data management and skeletons to build parallel applications,
% is the C++ Parallellism TS using SYCL to showcase ParallelSTL on CPU and GPU while the C++ TS only targets CPUs.
%% Proposals like GrPPI \cite{GrPPI:2017}, SkelCL \cite{SkelCL:2011} and SkePU \cite{SkePU:2010}
% and CUDPP \cite{}
%% provide composable primitives and skeletons to build parallel applications.
% CUDPP is CUDA-centered and it provides fixed operations.
%% GrPPI gives interesting reusable patterns for stream and data-parallel processing with many backends, but not OpenCL.
% SkePU and SkelCL
% They provide data management,
but the programmer is responsible of using their own data containers.

Works like VirtCL \cite{Virtcl}, MultiCL \cite{MultiCL} and SOCL \cite{SOCL} do address load balancing while abstracting the underlying system and managing data movement. Nevertheless, their focus is on task-parallelism instead of on the co-execution of a single data-parallel kernel.
%% a flexible  that ohigher\Cpp{} These \Cpp{} projects provide similar STL-based and lower-level API than the offered by EngineCL.

Also, there are C-programmed libraries with similar objectives, but they provide low-level APIs where the programmer needs to specify many parameters and the density of the code is considerable. While Maat \cite{Maat:2009} uses OpenCL to achieve the code portability, Multi-Controllers \cite{MultiControllers:2017} is CUDA and OpenMP-targeted.

On the other side, EngineCL is tightly coupled to OpenCL, offers a higher-level layered API with high usability, and its approach is to provide constructs that offer greater abstraction of the data-parallel problem while it has adaptiveness to different scenarios.
%% On the other side, EngineCL targets a % higher-level
%% flexible
%% API with an application domain as execution unit, increasing significantly the productivity. It provides different API layers, allows kernel specialisation, direct usage of \Cpp{} containers, manages the data and work distribution transparently between devices and has negligible overheads compared with the previous projects.

% NOZAL: no he metido SnuCL ni StarPU por el siguiente motivo:
% SkePU es el skeleton de StarPU, así que ya no menciono a StarPU.
% SnuCL es para nodos y clusters, el compilador hace transformación de código, tiene scheduling y mezcla MPI+OpenCL.
% Multi-Controllers: está hecha en C y usa OpenMP y CUDA (no está centrada en OpenCL), pero permite usar diversos dispositivos, especialización de kernels y balancear carga, pero la API es de más bajo nivel y el programador tiene que gestionar la memoria a nivel de controlador.

%% file: sections/conclusions.tex
%% \vspace{-5mm}
\section{Conclusions and Future Work}
\label{sec:Conclusions}
%% \vspace{-3mm}

%JLB: Modificar ligeramente cuando tengamos los números definitivos del equilibrio de carga.
Given the great relevance of heterogeneous systems in all sectors of computing, it is necessary to provide the community with tools that facilitate their programming without performance overheads. EngineCL is an OpenCL-based runtime that greatly simplifies the programming effort while frees the programmer from tasks that require a specific knowledge of the underlying architecture. Thanks to its load balancing capabilities, it can make use of all the available resources in the system, reducing the computation time.

% añadir la posibilidades d eoptimización para la coejecución
% reducir

This paper focuses on commodity systems and programs with small execution times, where it is a challenge to make co-execution cost-effective due to runtime overheads and the workload distribution. To overcome these difficulties two different approaches are researched. First, a number of optimizations have been designed and implemented on EngineCL that address the initialization and buffer management. Second, an effort has been made to optimize the HGuided scheduler, provided in EngineCL. For this purpose, an exhaustive experimental evaluation has been carried out to tune the values of the parameters of this algorithm under the worst-case scenario for load balancing.

A number of conclusions can be drawn from the experimental results. Firstly, the proposed optimizations reduce the execution time of a program by 7.5\% and 17.4\% to make co-execution successful for the binary and ROI operation modes, respectively. Regarding the HGuided parameters, it can be concluded that the more powerful the device, the larger the minimum packet size and the lower the k parameter value should be. Thus, the best result is always obtained if the minimum packet size is not limited for the CPU, when being the less powerful device in the system. Finally, thanks to all the optimizations, the new load balancing algorithm is always the fastest and most efficient scheduling configuration, yielding an average efficiency of 0.84. Thus, opportunities to compete in time-constrained scenarios against the fastest device increase.
% Acaban yielding en HGuided optimizado => superar al resto de configurationes/planif

%%JLB: Esto es previo.
%The API provided to the programmer is very simple, thus improving the usability of heterogeneous systems. This statement is corroborated by the exhaustive validation that is presented, with a large variety of Software Engineering metrics, achieving excellent results in all of them. On the other hand, the careful design and implementation of EngineCL allows zero overheads with respect to the native OpenCL version in some cases.
%% that in some of the experiments carried out, it obtains zero overheads with respect to the native OpenCL version.
%% allows that in many of the experiments carried out, it obtains slight improvements with respect to the native OpenCL version.
%In the rest of the cases, the overhead due to the management performed by EngineCL is negligible, below 3\% in all the cases studied and with an average overhead of 0.46\%, achieving an excellent portability performance. Also, two load balancing algorithms have been implemented and validated in order to give the best performance to both regular and irregular applications. The use of the whole heterogeneous system is always beneficial for at least one load balancing method, achieving an average efficiency of 0.84 when selecting the best scheduling configuration. This excellent performance, together with its proven usability, makes EngineCL a powerful tool for exploiting all kind of heterogeneous systems.

In the future, it is intended to extend the EngineCL to support iterative and multi-kernel executions, imitating the ROI operation mode of real applications. Also, new commodity architectures and load balancing algorithms will be provided and studied, focusing on performance and energy efficiency.
% to support a suitable co-execution on multiple devices simultaneously.
% Also, load balancing algorithms will be provided and studied as part of the scheduling system to support a suitable co-execution on multiple devices simultaneously.